\documentclass[prl,amsmath,aps,onecolumn]{revtex4}
\usepackage{graphicx,xspace}
\usepackage{float}
\usepackage{amsmath}
\usepackage{amssymb}
\usepackage[normalem]{ulem}
\input{epsf}
\begin{document}
\def\tit#1#2#3#4#5{{\sl #5} #1 {\bf #2} #3 (#4)}
\newcommand{\VEC}[1]{\mathbf{#1}} \newcommand{\TO}{,\ldots,}
\newcommand{\rvec}{\VEC{r}} \newcommand{\scalar}[2]{\left \langle#1\ 
    #2\right \rangle} \newcommand{\me}{\mathrm{e}}
\newcommand{\mi}{\mathrm{i}} \newcommand{\dif}{\mathrm{d}}
\newcommand{\fcs}{f_{\text{c}}^{\text{smpl}}}
\newcommand{\msmall}{\scriptscriptstyle}
\newcommand{\efes}{\msmall \text{FS}}
\newcommand{\bra}{\langle} 
\newcommand{\ket}{\rangle}
\newcommand{\half}{\frac{1}{2}}
\newcommand{\mean}[1]{\langle #1 \rangle}
\newcommand{\fig}[1]{figure~\ref{#1}}
\newcommand{\const}{\text{const}}
\newcommand{\elastic}{\text{elastic}}
\newcommand{\random}{\text{random}}
\newcommand{\eq}[1]{eq.~\ref{#1}} \newcommand{\gl}{\!=\!}
\newcommand{\pf}{\!\rightarrow\!}
\newcommand{\mq}[2]{\uwave{#1}\marginpar{#2}} 
\newcommand{\SET}[1]{\{#1\}}

\title{Depinning exponents of the driven long-range elastic string} 
\author{Olaf Duemmer} \email{duemmer@lps.ens.fr}
\author{Werner Krauth} \email{krauth@lps.ens.fr}
\affiliation{CNRS-Laboratoire de Physique Statistique \\
  Ecole Normale Sup{\'{e}}rieure\\
  24 rue Lhomond, 75231 Paris Cedex 05, France}
\begin{abstract} 
  We perform a high-precision calculation of the critical exponents
  for the long-range elastic string driven through quenched disorder at the
  depinning transition, at zero temperature. Large-scale simulations
  avoid finite-size effects and improve accuracy. We explicitly
  demonstrate the equivalence of fixed-velocity and fixed-driving-force
  simulations. The roughness, growth, and velocity exponents are
  calculated independently, and the dynamic and correlation length
  exponents are derived. The critical exponents satisfy known scaling relations
  and agree well with analytical predictions.
\end{abstract}

\maketitle

Driven elastic manifolds in disordered media  model a multitude of
physical systems ranging from cracks advancing in a solid in response
to an external strain field \cite{GaoRice} to charge density waves
\cite{Gruener}, vortices in type-II superconductors in an electric field
\cite{Blatter}, and to magnetic domain walls in an external magnetic field
\cite{Lemerle}.  These widely studied out-of-equilibrium systems undergo
dynamic phase transitions \cite{Fisher85}  as a function of the external
driving force (strain, electric or magnetic field, etc.).  At driving
forces below the depinning threshold ($f<f_c$), the disorder pins the
manifold so that its center-of-mass velocity vanishes at large times.
Above the depinning  threshold $f_c$, avalanches  advance the manifold
with finite center-of-mass velocity. At the depinning threshold (for
$f \rightarrow f_c^+$), the center-of-mass velocity in the long-time
limit goes to zero algebraically, whereas the typical length, width
and duration of the avalanches diverge. This critical divergence is
characterized by universal scaling exponents.

Most simply, advancing crack fronts are modeled by the dynamics of
an elastic string driven through quenched disorder (the inhomogeneous
material). On that level, elasticity theory combined with a reasonable
energy dissipation mechanism can be reduced to an effective equation of
motion for the crack front \cite{GaoRice,RamaErtasFisher,BBLP}. The
resulting elastic energy kernel is long ranged, i.e. falls of as
the squared inverse distance. The same long-range elastic energy
kernel is found in the wetting problem of an advancing contact line
\cite{JoannyDeGennes}.

In the last fifteen years, well-controlled experiments
have been designed to measure universal scaling exponents
\cite{Prevost,Moulinet,Schaffer,Santucci, PonsonMethod}. But the
experimental precision available does not yet allow to give definite
answers on questions about universality. In addition, experimental
results on the roughness exponent of contact lines are inconsistent
with theoretical predictions.  However, a long-standing controversy on
numerical values of the roughness exponent has recently been resolved
\cite{FisherFRG,Ertas,ChauveFRG,FisherAutomat,Tanguy,Schmittbuhl,RossoKrauth}.

In this paper, we perform a numerical simulation of the depinning phase
transition of the long-range string driven through quenched disorder, with
the aim of computing the critical exponents to high precision. Numerical
simulations of this model are non-trivial because of large finite-size
effects, caused by the unusually large value of the finite-size-scaling
exponent.

The discrete long-range elastic string is represented by a vector of
integer heights $\SET{h_1 \TO h_L}$ with periodic boundary conditions.
The string follows discrete quasi-static dynamics while being driven
through a discrete disorder (cf \cite{NattermannModel}).  The long-range
elastic force  acting on a string element $i$ is given by
\begin{equation}
  \label{eq_lin_el}
  f_i^\elastic = \sum_{j\ne i}^L \frac{h_j-h_i}{|j-i|^2}.
\end{equation}
Even though the sum contains $\sim L$ terms for each of $L$ sites, all
forces on all sites can be computed in $\sim L \log L$ operations using
a fast-Fourier-transform (FFT) algorithm.

The quenched disorder generates a random force $\eta$: 
\begin{equation}
  \label{eq_lin_ran}
  f_i^\random = f + g \eta_i(h_i).
\end{equation}
While $f$ is a constant external field, the random position-dependent
force $\eta$ is drawn from a binomial distribution: $\eta$ equals
plus one with probability $p$, and minus one with probability $1-p$.
The coupling $g$ \cite{g_is_one} controls the Larkin length at which
the elastic and disorder forces balance.

The random force results in a net external driving force of $f +
g(2p-1)$. The parameter $f$ can be adjusted to keep constant the
otherwise fluctuating center-of-mass velocity $v$ of the string. 
But we use fluctuating-velocity/fixed-force
simulations throughout this paper, unless stated otherwise. $p$ is the
control parameter of the dynamic phase transition: its distance to the
critical value $p_c$ (the depinning threshold) controls the large-scale
properties of the dynamics, the divergence of the correlation length and
the algebraic decay of the mean velocity $v$ \cite{Surprising}.  The skewed
binomial distribution of random forces corresponds to random-field frozen
disorder, as opposed to random-bond disorder.

The microscopic dynamics of the string is given by the following discrete dynamical rule:
\begin{equation}
  h_i(t+1)-h_i(t) = v_i(t) = 
  \begin{cases}
    1 & \text{if $ f_i^\elastic + f_i^\random > 0$}\\
    0 & \text{otherwise}
  \end{cases}\quad t=0,1,\dots\;.
  \label{velocity_definition}
\end{equation}
This rule does not allow for backward motion, as is justified by the
{\em no-return} theorem \cite{Middleton}.

Very large system sizes are crucial in order to calculate the critical
exponents with sufficient precision. Consider for example the velocity
exponent $\beta$, defined as $v \sim (p-p_c)^{\beta}$. For it to have an
accuracy of $\delta \beta = 0.01$ requires a resolution of $\delta p_c =
10^{-3}$ in the critical force \cite{delta_p_c}. This resolution can only
be obtained if the finite-sample-size fluctuations of the critical force
$\sigma_{p_c}$ are small enough. From a finite-size-scaling analysis one
finds $\sigma_{p_c}(L) = 0.65 \times L^{-1/\nu_{\efes}},\, \nu_{\efes}
= 1.62$ (data not shown), and demanding that $\sigma_{p_c} < 10^{-3}$
gives the minimum system size $L_{\text{min}} \approx 10^5$.

The discretized model, with its coarse and low-cost description of
microscopic dynamics, can be simulated for sizes up to $L=2^{20}\approx
10^{6}$, more than three orders of magnitude larger than for earlier discrete
\cite{Tanguy,Schmittbuhl} and continuum simulations \cite{Duemmer}.

To benefit from the high speed of the FFT algorithm, we use parallel
updating: The elastic force on all string elements is calculated. Then, 
the dynamic rule (equation (\ref{velocity_definition})) is applied
simultaneously to all sites. Finally, the random force is updated for those
sites that have stepped forward. In this way a single vector of length $L$
suffices to store the random force, avoiding storage problems even for
very large system sizes.  The parallel updating leads to a fluctuating
mean velocity $v$, whereas the external driving force is kept constant
via the fixed control parameter $p$.

The depinning threshold $p_c$ is determined by adapting a technique
known from absorbing-state dynamic phase transitions \cite{Hinrichsen}:
we start the simulation for fixed $p,\;f=0$ from a flat line at
$t=0$, and monitor the ratio $\Upsilon$ of string width $w(t) \equiv 
 [(1/L) \sum_i^L  (h_i - \bra h\ket)^2 ]^{1/2}$ to decaying center-of-mass velocity $v(t) \equiv 
(1/L) \sum_i^L v_i(t)$ times time $t$:
\begin{equation*}
\text{At threshold}:\quad
v(t) \sim \frac{w(t)}{t} = t^{\zeta/z -1},\quad 
          \implies \quad \Upsilon(t) \equiv \frac{w(t)}{t v(t)} \sim \text{constant}.
\label{}
\end{equation*}
\begin{center}
   \begin{figure}[t]
       \centerline{\includegraphics[bb=95 82 335 244]{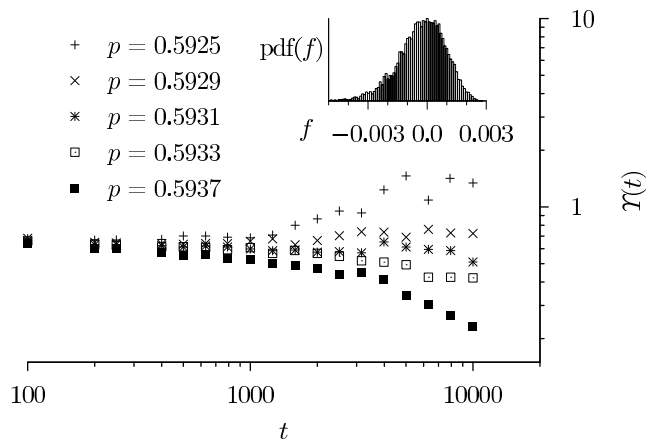}}
       \caption{At the depinning threshold $p_c$, the ratio  
       $\Upsilon(t)= w/[tv(t)]$ is constant in time, as
       shown by scaling arguments (for $p<p_c$, $\Upsilon(t \rightarrow
       \infty) \rightarrow \infty$ because the velocity vanishes, and for
       $p>p_c$, $\Upsilon(t \rightarrow \infty) \rightarrow 0$).
       The inset shows fluctuations around zero of the external 
       force $f$ for a simulation with
       fixed $p_c$ and fixed velocity: one percent of all sites move at each time step.
       }
      \label{fig_wvt_t}
   \end{figure}
\end{center} 
To understand why the velocity at threshold is given by the width divided
by time, we remark that, intuitively, an avalanche takes time $t$ to
advance a distance equal to its width $w(t)$. Moreover, at threshold, the
motion of the string is solely due to avalanches, hence the mean velocity
is given by $v(t)=w(t)/t$. It follows that the ratio $\Upsilon(t)$ is
constant at threshold. Away from threshold it either grows to infinity
(for $p < p_c$) or decays to zero (for $p>p_c$), providing a very
sensitive indicator for the depinning threshold.

\begin{figure}[b]
  \includegraphics[bb=93 80 325 244]{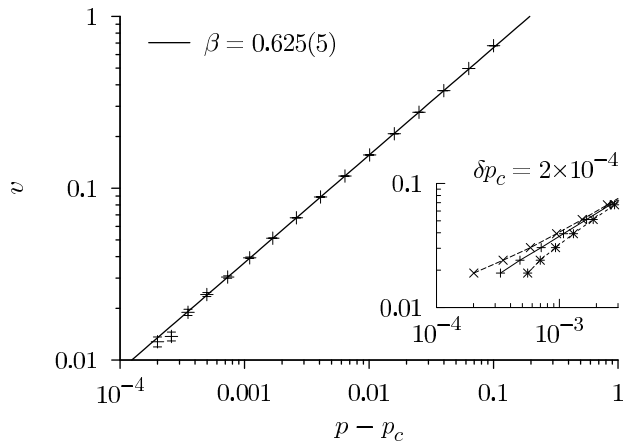}
  \caption{{\em Velocity exponent $\beta$.} The steady-state mean
    velocity as a function of the control parameter $v \sim
    (p-p_c)^{\beta}$ gives the velocity exponent: $\beta=0.625(5)$. Inset:
    slightly wrong values of $p_c$ lead to non-zero curvature
    ($(\times)\;p_c\rightarrow p_c-\delta p_c\;,\;(\ast)\;p_c\rightarrow
    p_c+\delta p_c$).
  }
  \label{fig_beta}
\end{figure}

The main plot in \fig{fig_wvt_t} illustrates the sensitivity of
$\Upsilon(t)$ to tiny changes in $p-p_c$, which in turn enables a very
precise localization of the critical threshold $p_c=0.5931(2)$. $p_c$
can alternatively be extracted from a power-law fit of the steady-state
mean velocity ($v \sim (p-p_c)^\beta$, see \fig{fig_beta}).

There has been debate about the equivalence between simulations
with fluctuating velocity/fixed force on the one hand and with fixed
velocity/fluctuating force on the other.  In the thermodynamic limit, one
expects them to be equal because the fluctuations of the center-of-mass
velocity vanish on large enough length scales.  In experimental or
numerical systems, the center-of-mass velocity fluctuates only on scales
smaller than the correlation length.  In our very large simulations, we
are able to check the equivalence between the two approaches: in addition
to the fluctuating-velocity simulations in the rest of this paper, we
perform a fixed-velocity simulation by adjusting the parameter $f$ at each
time step such that a very small but fixed proportion of sites move, thus
automatically tuning the system  to the depinning threshold.  The inset
of \fig{fig_wvt_t} shows the minute fluctuations of the external force
$f$ around a zero mean value in a simulation where the external force is
chosen at each time-step such that one percent of sites are made to move
\cite{why_v_0.01}, at $p=p_c$.  This demonstrates that $p_c$ is indeed
the critical point, and that the two types of simulations are equivalent.

\begin{center}
  \begin{figure}[t]
    \centerline{\includegraphics[width=8cm,bb=91 82 329 247]{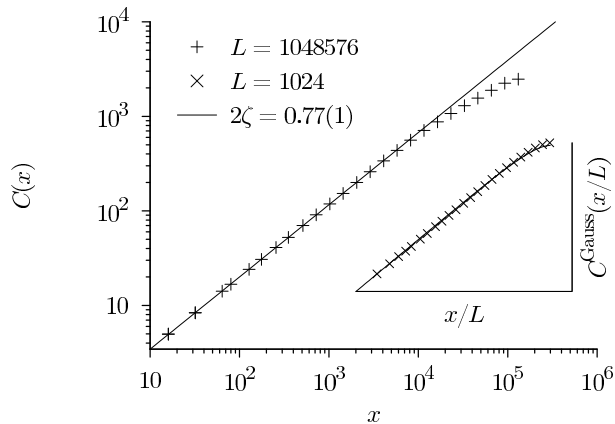}}
    \caption{{\em Roughness exponent $\zeta$.} At short distances,
      the two-point correlation function scales as 
      $C(x) \propto x^{2 \zeta}$. A linear fit gives $\zeta = 0.385(5)$
      (average taken over pinned configurations exactly at $p=p_c$).
      The inset compares correlation-function data  (for $L=1024$)
      with the Gaussian model of (\eq{fit_all_delta}).}
    \label{fig_zeta}
  \end{figure}
\end{center}
The roughness exponent $\zeta=0.388$ \cite{RossoKrauth} characterizes the
self-affine spatial structure of the string at depinning, and the scaling
properties of its average width $w$
(see also \cite{VDB}).  As a check for the discrete model, we compute this
exponent from the  short-distance behaviour of the two-point correlation
function $C(x) \equiv \mean{ (h(0)-h(x))^2 } \sim x^{2\zeta}$. Averaging
over pinned configurations exactly at the depinning threshold $p=p_c$,
we confirm earlier results ($\zeta=0.385(5)$, see figure \ref{fig_zeta}).

The correlation function $C(x)$ can be fitted over the whole interval
$x \in [0,0.5L]$ using the Gaussian model \cite{Rosso_Santa}: suppose
the random string configuration $h(x)$ is decomposed into Fourier modes

\begin{equation*}
   h(x) = \sum_n a_n \cos(2\pi n x/L) + b_n \sin(2 \pi n x/L), 
\end{equation*}
where $a_n$ and $b_n$ are Gaussian random variables with variance
$\sigma^2_n\propto 1/n^{1+2 \zeta}$. The resulting two-point correlation
function is given by $C^{\text{Gauss}}(x)= 4 \sum_n \sin^2(\pi n x/L)/n^{2
\zeta+1}$, with the following expansion for small $x/L$ \cite{Rosso_Santa}
(note the difference between the roughness exponent $\zeta$ and the
Riemann zeta-function $\zeta_{\text R}$):
\begin{multline}
  C^{\text{Gauss}}(x) = \text{constant} \bigg\{ - \left(\frac x L\right)^{2 \zeta} 
  2 ^{-1 + 2 \zeta}\pi^{2 \zeta} \Gamma(-2 \zeta) \sin\frac{\pi (2 \zeta + 1)}{2} \\
  + \left(\frac x L\right)^2 \frac{2 \pi^2}{2!} \zeta_{\text R}(2\zeta-1) 
  - \left(\frac x L\right)^4 \frac{8 \pi^4}{4!} \zeta_{\text R}(2\zeta-3) + O\left[\left(\frac x L\right)^6\right] \bigg\}.
  \label{fit_all_delta}
\end{multline}
This function is used in the inset of \fig{fig_zeta} to fit
the correlation function data obtained with the exact algorithm
\cite{AlbertoAlgo} for smaller systems. The fit remains excellent even
for large $x/L$.
\begin{center}
  \begin{figure}
    \centerline{\includegraphics[bb=91 82 329 244]{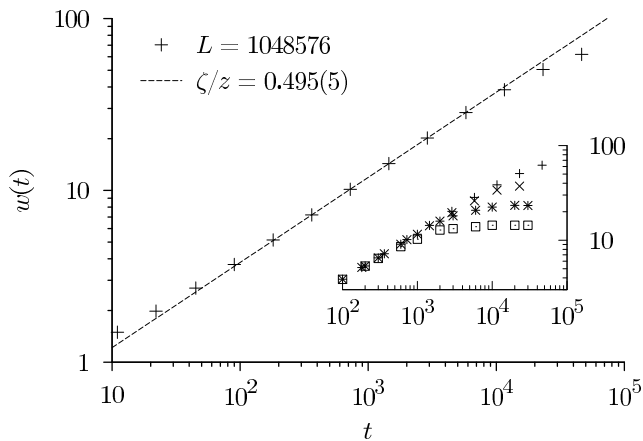}}
    \caption{{\em Growth exponent $\zeta/z$}. The algebraic growth of
    the average width $w(t)\sim t^{\zeta/z}$, at the depinning threshold
    $(p=p_c)$ and after initial relaxation effects $(t\sim 100)$, gives
    the growth exponent $\zeta/z=0.495(5)$. Inset: large finite-size
    effects as width levels off at $w_L \sim L^{\zeta}$, for $L =
    \SET{2^{14}\TO 2^{20}}$.}
    \label{fig_growth}
  \end{figure}
\end{center}
\newpage
The dynamic exponent $z$ was previously estimated for discrete extremal
models of limited size \cite{Tanguy,Schmittbuhl}. We determine it via the
growth exponent $\zeta/z$, which controls the growth of the average width
with time $w(t) \sim t^{\zeta/z}$. This algebraic growth law results
from the two scaling laws for the width and the time in terms of the
correlation length $\xi$: $w \sim \xi^{\zeta}$ and $\xi \sim t^{1/z}$.
We start at $t=0$ from a flat line, at the critical point $p = p_c$,
and monitor the growth of the average width. Figure \ref{fig_growth}
shows how, after an initial relaxation phase \cite{AlbertoRelax}, the
width grows algebraically with the growth exponent $\zeta / z = 0.495(5)$,
implying $z = 0.770(5)$ for the dynamic exponent.  The width $w(t)$ is
also affected by large finite-size effects. The algebraic growth regime
is limited by the width of the entire string $w_L \sim L^{\zeta}$. The
inset of figure \ref{fig_growth} demonstrates how $w(t)$ saturates at
the system-size-dependent limit $w_L$. The plots for different system
sizes $L$ can be superposed by plotting $w(t)L^{-\zeta}$ versus $tL^{-z}$
(not shown).

\begin{figure}[t]
  \includegraphics[bb=89 82 329 247]{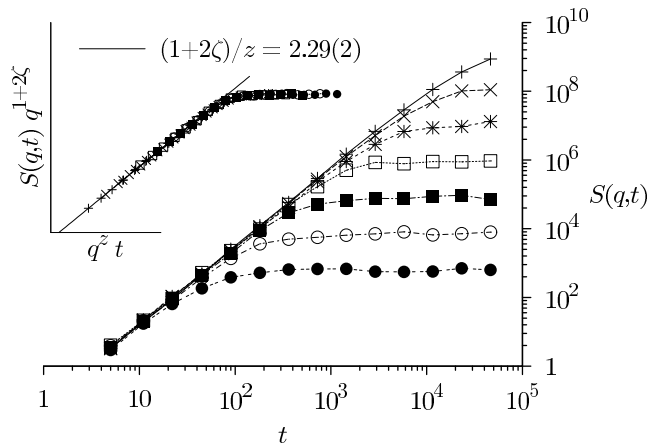}
  \caption{{\em Time-dependent structure factor $S(q,t)$.}
    Different Fourier components [from $(+)\; q=3\times10^{-6}$ to
    $(\bullet)\; q=10^{-2}$] take a relaxation time of $q^{-z}$ to reach
    their steady-state value of $q^{-1-2\zeta}$.  Inset: the small-time
    behaviour of the universal scaling function gives a further estimate
    of the dynamic exponent $z=0.77(1)$.
  } 
  \label{fig_sqt}
\end{figure}

Another way of determining the dynamic exponent $z$ is to analyze
the time-dependent structure factor $S(q,t)\equiv  \bra h(q,t)h(-q,t)
\ket$ with the Fourier transform of the height given by $h(q) \equiv
\sum_{n=0}^{L-1} \exp(\mathrm{i}q n) h_n$ and $q \equiv 2\pi k/L$,
where $k\in \SET{0 \TO L-1}$ is the mode number.  The structure factor
allows us to illustrate both the dynamic growth of the interface and its
self-affine structure in space-time \cite{Barabasi}:
\begin{equation}
  S(q,t) \sim q^{-1-2\zeta} \Phi(q^{z}t) \quad\text{with}\quad \Phi(y)=
  \begin{cases} 
    y^{\frac{1+2\zeta}z} \;,\; y \rightarrow 0 \\
    \text{constant} \;,\; y \rightarrow \infty
  \end{cases} .
\end{equation}
Figure \ref{fig_sqt} illustrates the convergence of the different
Fourier modes $S(q,t)$ to their steady-state value $\sim q^{-1-2\zeta}$
after a relaxation time $q^{-z}$.  The small-time behavior $S(q,t)
\sim t^{(1+2\zeta)/z}$ provides the dynamic exponent, and we confirm
the numerical value $z = 0.77(1)$, albeit with lower precision
than in our analysis using $w(t)$. The inset of \fig{fig_sqt} plots
$S(q,t)q^{1+2\zeta}$ versus the dimensionless quantity $q^z t$. It shows
the universal scaling form $\Phi$, and demonstrates the self-affine link
between time $t$ and (inverse) space $q$, via the dynamic exponent $z$.

The velocity exponent $\beta$ controls the power-law decrease of
the mean velocity with the distance to the critical threshold:
$v \sim (p-p_c)^{\beta}$.  It is obtained from the center-of-mass
velocity $v$ in the steady-state regime
(see \fig{fig_beta}).  From a fit at fixed $p_c$, we obtain $\beta =
0.625(5)$. The data of the steady-state velocity can also be fitted to
obtain the threshold $p_c$, as shown in the inset of \fig{fig_beta}. Only
for $p=p_c$ is there a straight line on the $\log - \log$ plot. For $p
\ne p_c$, a clear non-zero curvature emerges.

Note that it is quite difficult to obtain a precise estimate for $\beta$,
again because of large finite-size effects. Trying to calculate $\beta$
on medium-size systems one faces the problem of a significantly reduced
critical scaling regime. It manifests itself already in the short-range
elastic model \cite{Duemmer}, and is even more pronounced on long-range
elastic interfaces of intermediate size (data not shown). Our 
very large systems  allow to obtain satisfying accuracy.
\newpage
Our value for the velocity exponent yields an independent check of the
growth exponent $\zeta / z$, via two scaling relations that reduce the
set of four critical exponents to two independent ones: the statistical
tilt symmetry \cite{NattermannScaling} links the roughness exponent
to the correlation length exponent $\nu = 1/(1-\zeta)$, and a further
scaling relation links all four exponents $\beta = \nu (z - \zeta)$
\cite{NattermannScaling}. Using these two equalities and values of
$\zeta = 0.385(5)$ and $\beta = 0.625(5)$ we converge on a value for
the growth exponent of $\zeta / z = 0.500(5)$. These numerical values,
our final result, reflect our numerical measurements and the constraints
of the scaling relations, which are finally used to deduce the dynamic exponent $z =
0.770(5)$, and the correlation length exponent $\nu=1.625(10)$.

In conclusion, we have numerically studied the non-equilibrium depinning
phase transition of the long-range elastic string driven through
quenched disorder. The values obtained for the critical exponents
satisfy scaling relations and agree with analytical predictions from
functional RG \cite{FisherFRG,Ertas,ChauveFRG}.  For completeness
we recall the one- and two-loop ($\epsilon,\epsilon^2$) results:
$\zeta_{\epsilon} = 0.33,\;\zeta_{\epsilon^2}=0.47; \;\beta_{\epsilon}=
0.78,\;\beta_{\epsilon^2}=0.59;\;z_{\epsilon}=0.78,\;z_{\epsilon^2}=0.66;
\;\nu_{\epsilon} = 1.33,\;\nu_{\epsilon^2} =1.58$.  On the experimental
side, it will be interesting to understand why exactly the roughness and
growth exponents calculated in this paper agree with those measured in a
recent crack front experiment \cite{PonsonCeramics}, whether they can also
be seen in wetting experiments, and more generally, whether universality can
be confirmed as the physical mechanism unifying these diverse phenomena.

We thank D.~Bonamy, E.~Bouchaud, H.~Chat\'e, A.~Kolton, P.~Le~Doussal,
S.~Guibert, L.~Ponson, A.~Rosso, and K.~J.~Wiese for stimulating discussions.


\begin{thebibliography}{10}
\bibitem{Surprising} Surprisingly, setting $p=1/2$ while letting $f$
approach the critical value $f_c = g(2p_c-1)$ is not equivalent to our
procedure. The discreteness of the model entails in this case that the
asymptotic velocity does not approach zero continuously as $f \rightarrow
f_c$, it rather drops discontinuously from a value of order unity at $f >
f_c$ to zero at $f \le f_c$.
\bibitem{why_v_0.01} We choose $v=0.01$ because this is the limit of statistical resolution of a system of size $L\approx 10^6$ (cf figure \ref{fig_beta}). The sequential updating of a single site per time step leads to the smallest possible velocity and keeps the system exactly at the depinning threshold. Used in previous medium-size simulations \cite{Tanguy,Schmittbuhl}, it becomes however too slow for large systems, since all elastic forces ($\sim L \log  L$) have to be recalculated each time a single site advances.
\end{thebibliography}
\end{document}